\documentclass[pra,superscriptaddress,twocolumn]{revtex4}

\usepackage{enumerate}
\usepackage{amsfonts,amssymb,amsmath}
\usepackage[]{graphics,graphicx,epsfig}
\usepackage{amsthm}

\usepackage{graphicx}
\usepackage{dcolumn}
\usepackage{natbib}
\usepackage{color}
\usepackage{multirow}
\usepackage{ulem}

\begin{document}


\title{Synchronizing the simplest classical system and then quantizing it}

\author{Pawe{\l} Kurzy{\'n}ski}
\email{pawel.kurzynski@amu.edu.pl}
\affiliation{Faculty of Physics, Adam Mickiewicz University, Umultowska 85, 61-614 Pozna\'n, Poland}
\affiliation{Centre for Quantum Technologies,
National University of Singapore, 3 Science Drive 2, 117543 Singapore,
Singapore}

\date{\today}


\begin{abstract}
I propose a discrete synchronization model of finite d-level systems and discuss what happens once superposition of states is allowed. The model exhibits various asymptotic behaviors that depend on the initial state. In particular, two antagonistic phenomena can occur: a quantum-to-classical transition and entanglement generation. Next, I generalize this model and show that it is possible to phase-lock a periodic dynamics of a single qubit to a periodic dynamics of a qudit stimulus.
\end{abstract}

\maketitle


\section*{Introduction} 

Synchronization is a process in which two or more interacting oscillators adjust their rhythms. It can be found in many areas of classical physics  \cite{StrogatzSync} and was also studied in the quantum regime \cite{QS1,QS2,QS3}. Recently an interesting question was asked \cite{Qsync1}: {\it What is the smallest quantum system that can be synchronized?} The current answer to this question seems to be inconclusive, since there are arguments for and against single qubit synchronization \cite{Qsync1,Qsync2,Qsync3,Qsync4}. These contrary arguments originate from a problem of how to define a limit cycle in a quantum dynamics. Limit cycles are assumed to be prerequisites for synchronization \cite{Pikovsky}, but they can only emerge in nonlinear systems \cite{Strogatz}. However, quantum dynamics is fundamentally linear (apart from the measurement process), therefore it is crucial to gain a deeper understanding of how synchronization works in the quantum domain. 

Here I propose an alternative approach to the quantum synchronization problem. First, I start with a question: {\it What is the smallest classical system that can be synchronized?}  and then I ask: {\it How to quantize this system?} What is meant by {\it the smallest} is ambiguous. I choose an information-theoretic point of view, hence the smallest system is the one whose state is described by a minimal amount of information. Therefore, I focus on synchronization of d-level systems that are fully described by $\log d$ classical bits. My main motivation behind this approach comes from the fact that both, a classical d-level system and a qudit, can store the same amount of classical information. This fact allows one to compare the two systems from the computational efficiency perspective, especially in the context of any possible quantum information-processing advantage. Moreover, my approach is along the lines of research on complex dynamics of finite-state systems and pseudochaos \cite{KocarevSzczepanski,Lambic}. This line of research is particularly important since realistic computers, both classical and quantum, are fundamentally finite-state machines that cannot support the continuous nature of chaotic systems \cite{Lambic}. 
  
I am going to consider a simplified model in which I partially depart from the standard paradigms of synchronization \cite{Pikovsky}: (1) systems to be synchronized need to be self-sustained oscillators -- in the presence of dissipation they should be able to generate their own rhythms, hence the corresponding dynamical systems should have a limit cycle; (2) the interaction between oscillators is weak -- the oscillators should not get unified into a single oscillator; (3) the synchronization can be observed in a certain region of systems mismatch -- if the rhythm of one oscillator is changed the other oscillator should respond to this change. 

In particular, I am going to relax the paradigm (1). I will consider idealized conditions in which there is no dissipation of individual oscillatory motion, hence there is no need for a limit cycle. To justify my approach, let me stress that many synchronization models, such as Kuramoto model \cite{Kuramoto}, base on an elementary oscillator equation $\dot{\theta} = \omega$, where $\theta$ is the phase and $\omega$ is the frequency. This is the phase oscillator that has no amplitude variable. It allows one to focus on a more fundamental dissipative aspect of synchronization -- phase-locking, which cannot happen without existence of some attractor. Attractors do not exist in conservative systems, therefore some dissipation is needed to observe phase-locking. 

Due to the clear departure from the paradigm (1) some may argue that the model discussed in this work is not a true synchronization model. I do not mind and in fact I prefer to think of it as some kind of a toy model, or if one prefers -- pseudosynchronization. Nevertheless, this model offers one a possibility to focus on a mechanism behind the phase-locking and, more importantly, on how this mechanism works and what it offers once one goes quantum. In particular, I am going to show that, depending on an initial state of the oscillator and the stimulus, the phase-locking can lead to two antagonistic processes: quantum-to-classical transition and entanglement generation. Finally, I am going to focus on a phase-locking of the smallest quantum system -- a single qubit.


\section*{Simple synchronization model} 

Let me first recall a simple continuous classical model in which a single oscillator gets synchronized to a stimulus. A very nice pedagogical discussion of this model can be found in the chapter four of the book by Strogatz \cite{Strogatz}. Let $\theta\in [0,2\pi)$ be a phase variable of a stimulus that oscillates with a frequency $\Omega$
\begin{equation}\label{stimulus}
\dot{\theta}=\Omega.
\end{equation} 
Next, consider an oscillator whose frequency is $\omega$ (I assume $\omega \neq \Omega$) and whose phase variable is $\varphi \in [0,2\pi)$. The evolution of this oscillator is affected by the stimulus in the following way
\begin{equation}\label{oscillator}
\dot{\varphi} = \omega + K f(\theta-\varphi),
\end{equation} 
where $f(\theta-\varphi)$ is a periodic function that describes the way the oscillator responds to the stimulus and $K$ measures the strength of this response. One often takes $f(\theta-\varphi)=\sin(\theta-\varphi)$ which implies the following behaviour: the oscillator speeds up if $\varphi$ is behind $\theta$ and slows down if $\varphi$ is ahead of $\theta$. It is useful to define the phase difference $\Delta=\theta-\varphi$ whose evolution is given by 
\begin{equation}\label{delta}
\dot{\Delta} = \Gamma - K f(\Delta),
\end{equation} 
where $\Gamma = \Omega - \omega$. The phenomenon of phase-locking occurs once the phase difference settles down to some constant value $\Delta^{\ast}$ (here $\ast$ denotes a fixed point, not a complex conjugation), which implies $\Gamma = K f(\Delta^{\ast})$. This equation can be satisfied if $\Gamma/K$ is in the image of $f(\Delta)$, which defines the range of entrainment. If $\Gamma/K$ is outside of this range, the oscillator is not able to adjust its rhythm to the stimulus.


\section*{Discrete synchronization model}
 
In this section I am going to discretize Eqs. (\ref{stimulus}-\ref{delta}). At this point some readers may think of the circle map
\begin{equation}\label{circlemap}
\alpha_{t+1}=\alpha_t + \Gamma - \frac{K}{2\pi}\sin(2\pi \alpha_t),
\end{equation}
which is a particular time-discrete version of (\ref{delta}). However, the circle map is not what I am looking for since the variable $\alpha_t$ that describes the state of the system is continuous. The goal is to further discretize the circle map, i.e., to discretize the set of system's states. 

Let me assume that the phases of the stimulus and the oscillator are discrete and can take $d$ different values $\theta_t,\varphi_t \in S = \{0,1,2,3,\ldots,d-1\}$. The equations (\ref{stimulus}-\ref{delta}) become
\begin{eqnarray}
\theta_{t+1} &=& \theta_t + \Omega, \label{dstimulus}\\
\varphi_{t+1} &=& \varphi_t + \omega + G(\theta_t - \varphi_t), \label{doscillator}\\
\Delta_{t+1} &=& \Delta_t + \Gamma -  G(\Delta_t), \label{ddelta}
\end{eqnarray}
where $\Delta_t = \theta_t - \varphi_t$, $G(\theta_t-\varphi_t)$ is an analogue of $Kf(\theta_t-\varphi_t)$ and I assume that all values are taken mod $d$. Note, that $\theta_t,\varphi_t \in S$ for all $t$ which implies that $\Omega$ and $\omega + G(\theta_t - \varphi_t)$ must be in $S$ too. Since the choice of $\omega$ should not depend on  $G(\theta_t - \varphi_t)$ and vice versa, I assume that both of them are integers from $S$.

The key ingredient of synchronization is the function $G(\Delta_t)$. Phase-locking means that after some number of steps, say $\tau$, the phase difference $\Delta_t$ is fixed, i.e., $\Delta_{t} = \Delta^{\ast}$ for $t \geq \tau$. This implies $G(\Delta^{\ast})=\Gamma$, hence a prerequisite for phase-locking is that $\Gamma$ is one of the outcomes of this function. To include the notion of how strong the oscillator responds to the stimulus let me assume that the image of $G(\Delta_t)$ is in the set $E_K=\{-K,-K+1,\ldots,K\}$ (equivalently $E_K=\{0,\ldots,K,d-K,\ldots,d-1\}$), where $K<\frac{d}{2}$. This set defines an entrainment range because phase-locking is possible if $\Gamma \in E_K$. 

Let me choose 
\begin{equation}\label{G}
G_K(\Delta_t) = \left\{\begin{matrix}
\Delta_t ~~\text{if}~~ \Delta_t \in E_K, \\
0~~~\text{else}.~~~~~~~~ 
\end{matrix}\right.
\end{equation}
This function allows the oscillator to respond linearly to the stimulus if the phase difference is not larger than $K$. If the phase difference is larger than $K$ the oscillator does not respond at all and follows its own rhythm. Below I consider the asymptotic behaviour of the system for various choices of $\Gamma \neq 0$ and $\Delta_0$. 


First, let me consider the case $\Gamma\in E_K$. If $\Delta_0 \in E_K$ then $G_K(\Delta_0)=\Delta_0$ and $\Delta_1 = \Gamma$. But $\Delta_2 = \Gamma$ as well, therefore $\Delta^{\ast}=\Gamma$ and the system gets phase-locked in one step. On the other hand, if $\Delta_0 \notin E_K$ then $G_K(\Delta_0)=0$ and $\Delta_1 = \Delta_0 + \Gamma$. If $\Delta_1 \notin E_K$ then $\Delta_2 = \Delta_0 + 2\Gamma$. In general if for all $t\leq \tau$  $\Delta_t \notin E_K$ then $\Delta_{t}=\Delta_0 + t\Gamma$. However, there must be some $\tau$ ($\tau<d$) for which $\Delta_{\tau}\in E_k$. This is because $\Delta_t$ increases by $\Gamma$ and at some point the phase difference must fall inside $E_K$. Note that by assumption $|\Gamma|\leq K$, hence it is not possible to {\it jump over} $E_K$ while making a step $\Delta_t \rightarrow \Delta_{t+1}$. Therefore, $\Delta_t = \Delta^{\ast}=\Gamma$ for all $t>\tau$ -- the system gets phase-locket in $\tau$ steps. Note, that after the phase-locking the evolution of the oscillator is given by 
\begin{equation}\label{aspl}
\varphi_{\tau + t} = \theta_{\tau+t} - \Gamma.
\end{equation}

If $\Gamma \notin E_K$ then there are no fixed points. To prove it, let me assume that a fixed point $\Delta^{\ast}$ exists. This and (\ref{ddelta}) imply $G_K(\Delta^{\ast})=\Gamma$, but by definition the values of $G_K(\Delta_t) \in E_K$, which contradicts the initial assumption that $\Gamma \notin E_K$. In this case there is no phase-locking, but instead there is a {\it phase-drift} -- the phase difference constantly changes. Note, that due to finiteness of state-space the changes of the phase difference are periodic. 




\section*{Quantum synchronization model}

The next step is to quantize the above discrete model. To do so, let me first reformulate it in a vector space formalism. Instead of writing $\theta_t$ and $\varphi_t$ I will write $|\psi_t\rangle = |\theta_t\rangle\otimes|\varphi_t\rangle \in \mathbb{C}^d\otimes\mathbb{C}^d$. Let me define an orthonormal basis $\{|i\rangle \}_{i=0}^{d-1}$ and let me assume that both $|\theta_t\rangle$ and $|\varphi_t\rangle$ are some vectors from this basis. Let me also define two hermitian operators
\begin{equation}\label{opthetaphi}
\hat{\theta} = \left(\sum_{i=0}^{d-1}i|i\rangle\langle i|\right)\otimes \hat{\openone},~~~~\hat{\varphi} =\hat{\openone} \otimes \left(\sum_{i=0}^{d-1}i|i\rangle\langle i|\right).
\end{equation}
These operators allow one to recover $\theta_t = \langle \psi_t|\hat{\theta}|\psi_t\rangle$ and $\varphi_t = \langle \psi_t|\hat{\varphi}|\psi_t\rangle$.

Next, let me define unitary operators $\hat{U}_{\Omega}$ and $\hat{U}_{\omega}$
\begin{equation}
\hat{U}_{\Omega}|\theta_t\rangle = |\theta_t+\Omega\rangle,~~~~\hat{U}_{\omega}|\varphi_t\rangle = |\varphi_t+\omega\rangle, \label{UOo}
\end{equation}
where mod $d$ convention is used. They describe the free evolution during which the stimulus and the oscillator follow their own rhythms. Eventually, let me define an operator $\hat{G}_K$ that acts on both systems
\begin{equation}\label{Ghat}
\hat{G}_K|\theta_t\rangle\otimes|\varphi_t\rangle =  \left\{\begin{matrix}
|\theta_t\rangle \otimes |\theta_t\rangle  ~~\text{if}~~ \Delta_t \in E_K, \\
|\theta_t\rangle \otimes |\varphi_t\rangle ~\text{else}.~~~~~~~~~~~
\end{matrix}\right.
\end{equation}
whose action mimics the one of (\ref{G}). Note that $\hat{G}_K$ is not unitary, however together with $\hat{U}_{\Omega}$ and $\hat{U}_{\omega}$ it allows one to iterate the state of the system in the following way
\begin{equation}\label{quantevol}
|\psi_{t+1}\rangle = (\hat{U}_{\Omega}\otimes \hat{U}_{\omega}) \hat{G}_K |\psi_t\rangle.
\end{equation} 

The lack of unitarity of $\hat{G}_K$ is related to its irreversibility. Note that this operator changes $2K+1$ different inputs into the same output $|\theta_t\rangle\otimes|\theta_t\rangle$. Such a transformation can be realized via Kraus operators. Equivalently, it can be realized unitarily by adding an ancillary system. I am going to follow the second approach.

I assume that the ancillary system has $d+1$ levels with the corresponding basis vectors $\{|\bar{0}\rangle\}\cup\{|i\rangle \}_{i=0}^{d-1}$, where $|\bar{0}\rangle$ is the initial state. Let me introduce a unitary operator $\hat{V}_K$ whose action is given by
\begin{equation}\label{V}
\hat{V}_K |\theta_t\rangle \otimes |\varphi_t\rangle \otimes |\bar{0}\rangle =  \left\{\begin{matrix}
|\theta_t\rangle \otimes |\theta_t\rangle \otimes |\bar{0}\rangle ~~~~~\text{if}~~\Delta_t=0,~~~~~~~ \\
|\theta_t\rangle \otimes |\varphi_t\rangle \otimes |\bar{0}\rangle ~~~~\text{if}~~ \Delta_t \notin E_K,~~~~ \\
|\theta_t\rangle \otimes |\theta_t\rangle \otimes |\Delta_t\rangle ~~~\text{else}.~~~~~~~~~~~~~~~~ 
\end{matrix}\right.
\end{equation}  
The operator $\hat{V}_K$ implements $\hat{G}_K$ on the first two systems. Moreover, note that the choice of $\hat{V}_K$ is non-unique. The reason why I chose it as (\ref{V}) is going to be explained later. Next, I impose that the ancilla is reset to $|\bar{0}\rangle$ after each application of $\hat{V}_K$ so that one can iterate the evolution more than once. This resetting is a source of dissipation that makes phase-locking possible. The above leads to the following formulation of Eqs. (\ref{dstimulus}-\ref{ddelta})
\begin{eqnarray}
\theta_{t+1} &=& \text{Tr}\{\rho_{t+1}\hat{\theta}\}, \label{rhostimulus} \\
\varphi_{t+1} &=& \text{Tr}\{\rho_{t+1}\hat{\varphi}\}, \label{rhooscillator} \\
\Delta_{t+1} &=& \text{Tr}\{\rho_{t+1}(\hat{\theta}-\hat{\varphi})\}, \label{rhodelta}
\end{eqnarray}
where
\begin{equation}\label{rhoevol}
\rho_{t+1} =\text{Tr}_{anc}\left\lbrace \hat{\mathbb{U}}  (\rho_t \otimes |\bar{0}\rangle\langle \bar{0}|_{anc}) \hat{\mathbb{U}}^{\dagger} \right\rbrace,
\end{equation} 
and
\begin{equation}\label{totalU}
\hat{\mathbb{U}}=(\hat{U}_{\Omega}\otimes \hat{U}_{\omega}\otimes \hat{\openone}) \hat{V}_K
\end{equation}

Up to now the system's state was assumed to be a product of two basis vectors, hence the dynamics was classical and one could choose either (\ref{dstimulus}-\ref{ddelta}) or (\ref{rhostimulus}-\ref{rhodelta}) to describe it. However, the latter set of equations allows one to consider a much bigger set of initial states. Below I am going to focus on the case $\Gamma\in E_K$ and discuss what happens if the initial state is a superposition of basis vectors.


{\it Synchronization to classical stimulus.} Let me first consider the case in which the stimulus is classical, i.e., it is prepared in one of the basis states, whereas the oscillator is prepared in a superposition. The initial state is $\rho_0 = |\psi_0\rangle\langle\psi_0|$, where
\begin{equation}\label{initial}
|\psi_0\rangle = |\theta_0\rangle \otimes \left(\sum_{i=0}^{d-1}\alpha_{i}|i\rangle\right).
\end{equation} 

The asymptotic behaviour of the system stems from (\ref{aspl}). The phase-locking of all terms occurs after $\tau$ steps and for $t>\tau$ the system's state is described by $\rho_{t} = |\psi_{t}\rangle\langle\psi_{t}|$, where $|\psi_{t}\rangle = |\theta_{t}\rangle \otimes |\theta_{t} -\Gamma \rangle$ and $\theta_{t}=\theta_0+\Omega t$. This state is a product of two basis vectors. It does not depend on the initial state of the oscillator. 

In simple words, the above scenario represents a quantum-to-classical transition -- the superposed oscillator gets phase-locked to the classical stimulus and becomes classical too. Note, that this transition occurs for any choice of $\hat{V}_K$ in (\ref{V}) that realizes $\hat{G}_{K}$.

{\it Synchronization to quantum stimulus.} Next, I consider a situation in which both, the stimulus and the oscillator, are prepared in a superposition of basis states. I am going to show that this state evolves into an entangled asymptotic state. Consider $\rho_0 = |\psi_0\rangle\langle\psi_0|$, where  
\begin{equation}\label{initial3}
|\psi_0\rangle = \left(\sum_{i=0}^{d-1}\alpha_i|i\rangle\right) \otimes \left(\sum_{j=0}^{d-1}\beta_j|j\rangle\right).
\end{equation} 

$\Gamma \in E_K$ implies that all terms from (\ref{initial3}) get phase-locked after some time $\tau$, hence the asymptotic state $\rho_{t}$ (for $t>\tau$) must be in the subspace ${\mathcal S}_{pl}$ spanned by $\{|i\rangle\otimes|i-\Gamma\rangle\}_{i=0}^{d-1}$ -- {\it the phase-locked subspace}. The only problem is to determine if $\rho_t$ has any coherences, i.e., non-zero off-diagonal terms. If the answer is positive, the partial transposition will move these terms outside of ${\mathcal S}_{pl}$ and the operator $\rho_{t}^{T_O}$ will be of the form $\sigma_{pl} \oplus \bar{\sigma}_{pl}$, where $T_O$ denotes partial transposition (transposition in the oscillator subspace). The part $\sigma_{pl}$ is inside ${\mathcal S}_{pl}$, whereas $\bar{\sigma}_{pl}$ is outside of ${\mathcal S}_{pl}$. Moreover, $\bar{\sigma}_{pl}$ is traceless (since there are no diagonal terms outside of ${\mathcal S}_{pl}$), hermitian and non-zero. This implies that $\rho^{T_O}_{t}$ has negative eigenvalues, thus $\rho_{t}$ is entangled \cite{PT1,PT2}. 

To prove that $\rho_t$ has some non-zero coherences, let me focus on the coherence between $|i+\Omega t\rangle\otimes |i+\Omega t-\Gamma\rangle$ and $|i'+\Omega t\rangle\otimes |i'+\Omega t-\Gamma\rangle$. The crucial observation in this proof is that any initial coherence in (\ref{initial3}) can only survive between the terms corresponding to the same initial phase difference. If the initial phase difference of one term is $\Delta$ and of the other is $\Delta'\neq \Delta$, then during phase-locking each term causes a different change of the ancilla's state and after tracing the ancilla the coherence is lost. As a result, the only initial coherences from (\ref{initial3}) that contribute to the examined one stem from 
\begin{equation}
|\Delta_0\rangle = \alpha_i\beta_{i-\Delta_0}|i\rangle \otimes |i-\Delta_0\rangle +\alpha_{i'}\beta_{i'-\Delta_0}|i'\rangle \otimes |i'-\Delta_0\rangle,
\end{equation}
where $\Delta_0 = 0,1,\ldots,d-1$. After phase-locking these vectors have the form
\begin{eqnarray}
|\Delta_{0,t}\rangle &=& \alpha_i\beta_{i-\Delta_0}|i+\Omega t\rangle \otimes |i+\Omega t-\Gamma\rangle  \\
&+&\alpha_{i'}\beta_{i'-\Delta_0}|i'+\Omega t\rangle \otimes |i'+\Omega t-\Gamma\rangle. \nonumber
\end{eqnarray}
The examined coherence results from a mixture of these terms and is of the form
\begin{equation}
\alpha_i\alpha^{\ast}_{i'}\sum_{\Delta_0 = 0}^{d-1}\beta_{i-\Delta_0}\beta^{\ast}_{i'-\Delta_0}.
\end{equation}
This value vanishes only for a set of amplitudes of measure zero and is maximal for $|\alpha_i|=|\beta_{i'}|=\frac{1}{\sqrt{d}}$. 

To sum up, I have shown that the superposed oscillator gets phase-locked to the superposed stimulus and as a result both systems get entangled. Interestingly, the entanglement is generated in a dissipative process. Moreover, unlike in the previous case, the outcome of the dynamics depends on a particular choice of $\hat{V}_K$ in (\ref{V}). More precisely, $\hat{V}_K$ was chosen such that it does not affect coherences in the phase-locked subspace ${\mathcal S}_{pl}$.


\subsection*{Synchronization of a single qubit}

The above model and all of the above examples work for an arbitrary $d$. In particular, for $d = 2$ the system reduces to a pair of qubits and all the parameters and variables are either zero or one. In this case a crucial limitation stems from binarisation of $K$: either each system evolves independently, or the oscillator gets completely slaved to the stimulus (this implies departure from the paradigm (2) -- see introduction). Still, the system is capable of phase-locking and of demonstrating a quantum-to-classical transition and entanglement generation. Nevertheless, the dynamics of such a two-qubit system is relatively simple, therefore in order to show that a single qubit can be synchronized in a more complex way I am going to generalize the previous model and will consider a synchronization of a qubit to a classical multi-level stimulus.

\begin{figure*}[t]
	\centering
\includegraphics[width=5.5cm]{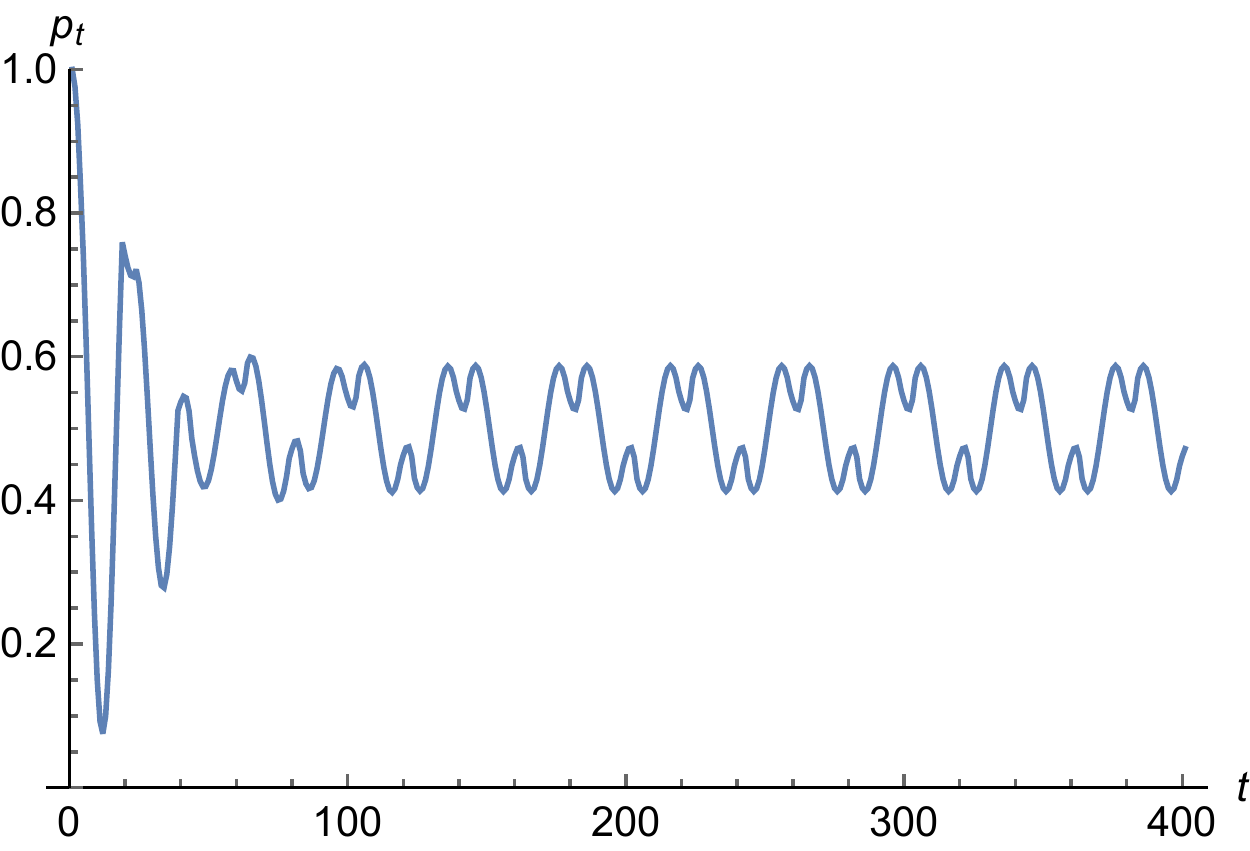}~~~~\includegraphics[width=5.5cm]{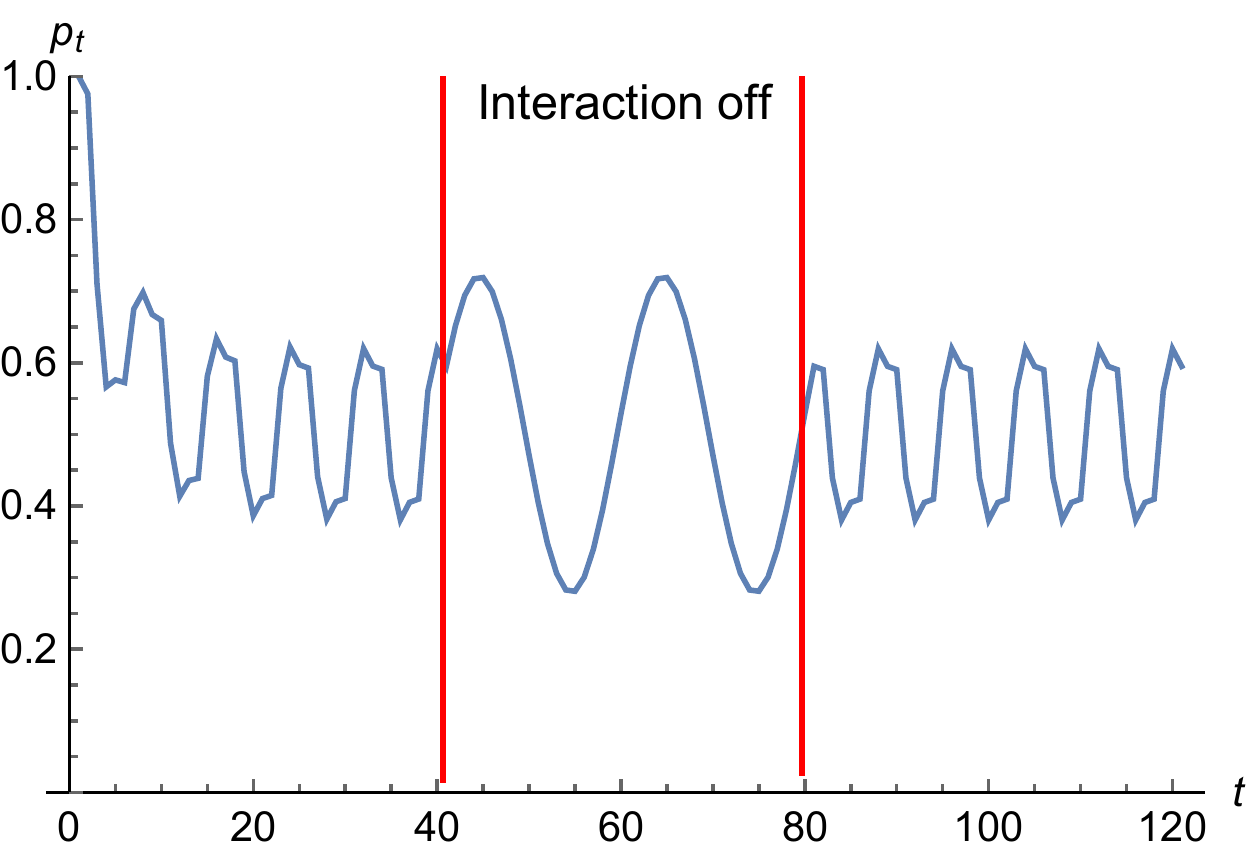}~~~~\includegraphics[width=5.5cm]{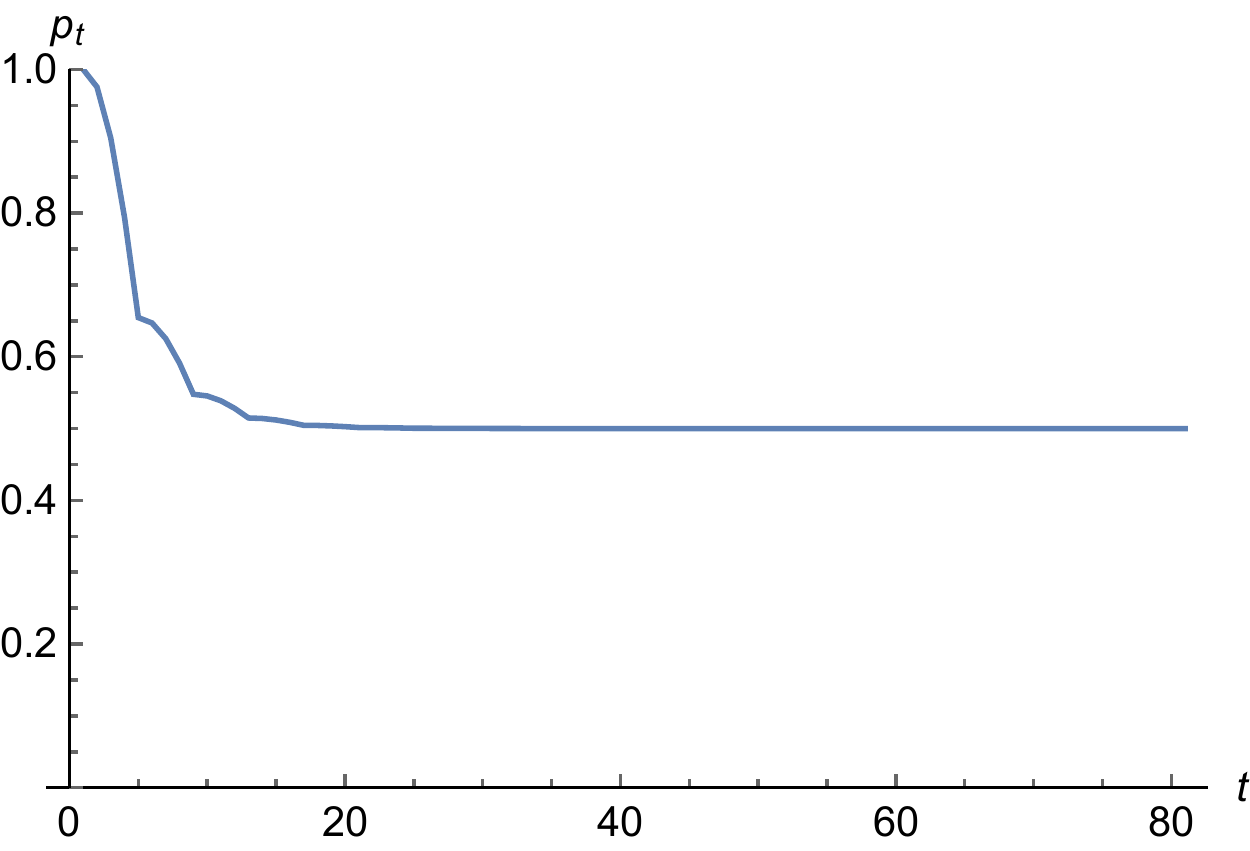}
	\caption{Numerical simulation of $p_t = \langle 0|\sigma_t|0 \rangle$ for  $d=40$. For a clarity of presentation points on the plots were connected. The single-qubit dynamics is described by Eq. (\ref{sigmaevol})  and the initial state is $\sigma_0 = |0\rangle\langle 0|$. Left ($\Omega = 1$, $\omega=2$, $K=2$): phase-locking, the qubit oscillates with the frequency $\Omega=1$ and the period of oscillations is $\frac{d}{\Omega}=40$. Middle ($\Omega = 5$, $\omega=2$, $K=5$): the qubit is phase-locked for steps $0-40$, next the interaction is turned off for steps $41-80$ and the qubit returns to its own frequency, then the interaction is turned on for steps $81-120$ and the qubit becomes phase-locked again. Right ($\Omega = 5$, $\omega=2$, $K=3$): the qubit depolarizes to a maximally mixed state.}
\label{fig1}
\end{figure*}

Let me assume that the stimulus is a d-level system, $\theta_t  = 0,1,\ldots,d - 1$, where $d$ is even, and the oscillator is a qubit, $\varphi_t = 0,1$. The free dynamics of the stimulus is going to be the same as in the previous case, but for simplicity I fix $|\theta_0\rangle = |0\rangle$, hence $|\theta_t\rangle = \hat{U}_{\Omega}^t|0\rangle = |\Omega t\rangle$. Moreover, $\Omega$ is chosen to be a divisor of $d$. For example, if $d=12$ then $\Omega = 1,2,3,4,6,12$. Such a choice guarantees that after $d$ steps the stimulus returns to $\theta_d = 0$. On the other hand, the dynamics of the oscillator is going to be determined by a unitary transformation
\begin{eqnarray}
\hat{R}|0\rangle &=& \cos \frac{\pi}{d} |0\rangle + \sin \frac{\pi}{d} |1\rangle, \\
\hat{R}|1\rangle &=& -\sin \frac{\pi}{d} |0\rangle + \cos \frac{\pi}{d} |1\rangle.
\end{eqnarray}
This transformation can be visualized using the Bloch-sphere picture as a $\frac{2\pi}{d}$-rotation about Y -axis. The natural frequency of the oscillator is given by $\omega$, which is also a divisor of $d$. The single step of the oscillator's free evolution is generated by $\hat{R}^{\omega}$, which corresponds to a $\frac{2\pi\omega}{d}$-rotation about Y-axis. It is also convenient to introduce the following qubit states $|\chi_{k}\rangle = \hat{R}^k|0\rangle$, where $k=0,1,\ldots,d-1$. 

Next, let me introduce a new interaction operator that requires an ancillary qubit (initially prepared in the state $|0\rangle$)
\begin{equation}\label{W}
\hat{W}_K |\theta_t\rangle \otimes |\varphi_t\rangle \otimes |0\rangle =  \left\{\begin{matrix}
|\theta_t\rangle \otimes |\varphi_t\rangle \otimes |0\rangle ~~\text{if}~~ |\theta_t - \frac{d}{2}\varphi_t|>K \\
|\theta_t\rangle \otimes |\chi_{\theta_t}\rangle \otimes |1\rangle  ~\text{else},~~~~~~~~~~~~~~~~~~ 
\end{matrix}\right.
\end{equation} 
and after each application the state of the ancilla is reset to $|0\rangle$. In simple words, $\hat{W}_K$ changes the state of the oscillator qubit to $|\chi_{\theta_t}\rangle$ if $|\theta_t-\frac{d}{2}\varphi_t|\leq K$. Let me recall that all values are taken mod $d$, as in the previous case. The multiplication of $\varphi_t$ by $\frac{d}{2}$ relabels the states of the qubit $\{0,1\} \rightarrow \{0,\frac{d}{2}\}$, so that they can be compared with the states of the stimulus. 

Intuitively, the stimulus can be considered as a classical hand of a clock that can point to one of $d$ different positions on a clock face. On the other hand, the oscillator qubit is a quantum hand that can point either vertically up ($\frac{d}{2}\varphi_t = 0$) or vertically down ($\frac{d}{2}\varphi_t = \frac{d}{2}$), or is in a superposition of these two possibilities. In the Bloch-sphere picture the state of the oscillator qubit is a vector that lies in the XZ-plane and the operator $\hat{W}_K$ tries to make its evolution follow the evolution of the stimulus hand on the clock face. 

A single step of the evolution is given by
\begin{equation}\label{rhoevol2}
\rho_{t+1} =\text{Tr}_{anc}\left\lbrace \hat{\mathbb{Q}}  (\rho_t \otimes |0\rangle\langle 0|_{anc}) \hat{\mathbb{Q}}^{\dagger} \right\rbrace,
\end{equation} 
where
\begin{equation}\label{totalU2}
\hat{\mathbb{Q}}=(\hat{U}_{\Omega}\otimes \hat{R}^{\omega}\otimes \hat{\openone}) \hat{W}_K,
\end{equation}
and $\rho_0 = |0\rangle\langle 0| \otimes \sigma_0$ is an initial state of the stimulus and the oscillator. This implies that $\rho_t = |t\rangle\langle t| \otimes \sigma_t$ and the goal is to understand the evolution of $\sigma_t$, whose general form is 
\begin{equation}
\sigma_t = p_t|0\rangle\langle 0| + \bar{p_t}|1\rangle\langle 1| + c_t|0\rangle\langle 1| + c_t^{\ast}|1\rangle\langle 0|,
\end{equation}
where $0 \leq p_t \leq 1$, $\bar{p_t}=1-p_t$ and $|c_t|^2 \leq p_t\bar{p_t}$.

The action of $\hat{\mathbb{Q}}$ can change $\sigma_t$ in three different ways 
\begin{equation}\label{sigmaevol}
\sigma_{t+1} =  \left\{\begin{matrix}
p_t \Pi_{\Omega t + \omega} + \bar{p_t} \Pi_{\frac{d}{2} + \omega}~~~\text{if}~~|\Omega t|\leq K, ~~~~~~~~\\
p_t \Pi_{\omega} + \bar{p_t} \Pi_{\Omega t  + \omega} ~~~~~~~\text{if}~~|\Omega t-\frac{d}{2}|\leq K,~~\\
\hat{R}^{\omega} \sigma_t \hat{R}^{\omega\dagger}~~~~~~~~~~~~~~~~\text{else},~~~~~~~~~~~~~~~~~~
\end{matrix}\right.
\end{equation} 
where $\Pi_{k}=|\chi_k\rangle\langle \chi_k|$. In the first two cases the coherences corresponding to $c_t$ and $c_t^{\ast}$ vanish due to the trace of the ancillary system. On the other hand, in the last case the ancilla remains uncoupled and the oscillator evolves unitarily. 

I am going to focus on how the probability $p_t = \langle 0|\sigma_t|0 \rangle$ changes in time. The formula (\ref{sigmaevol}) implies that $p_{t+1}$ can be derived from $p_t$ and $c_t$ via non-homogenous recurrence equation with variable coefficients. It is hard to follow it analytically, therefore I studied it numerically and the most important results are shown in Fig. \ref{fig1}. 

In order to minimize the chance of resonances between the stimulus and the oscillator I chose $\Omega$ and $\omega$ to be co-prime. The numerical studies show that for such a choice of frequencies there are essentially two types of behavior. The first behavior corresponds to the situation in which there exists $t$ such that $|\Omega t| \leq K$ or  $|\frac{d}{2} - \Omega t| \leq K$. In this case the oscillator qubit gets phase-locked, i.e.,  $p_t$ oscillates with frequency $\Omega$ and the period of oscillations is $\frac{d}{\Omega}$. If the interaction between the qubit and the stimulus is turned off, the qubit returns to its own frequency,  but it becomes phase-locked again once the interaction is turned on (see Fig. \ref{fig1}, middle). The second behavior corresponds to any other choice of $K$ and $\Omega$. In this case the oscillator qubit gets depolarized and evolves towards a maximally mixed state. There is a simple explanation for the second type of behavior. In such situation the only nontrivial application of $\hat{W}_K$ occurs for $\theta_t = 0$ (and $\theta_t = \frac{d}{2}$, provided there exists $t$ such that $\Omega t = \frac{d}{2}$). For all remaining $\theta_t$ the oscillator qubit evolves unitarily $\sigma_{t+1}=\hat{R}^{\omega} \sigma_t \hat{R}^{\omega\dagger}$. Note, that for $\theta_t = 0$ and $\theta_t = \frac{d}{2}$ the operator $\hat{W}_K$ causes an effective measurement of the oscillator qubit in the $\{|0\rangle, |1\rangle\}$ basis. Therefore, the unitary rotation about Y-axis is periodically interrupted by a dephasing along Z-axis, hence the length of the qubit's Bloch vector decreases and in the end the qubit becomes maximally mixed.


\section*{Concluding remarks}

I considered a synchronization model in which both, time and state-space are discrete. The phase-locking of the oscillator to the stimulus occurs due to an irreversible transformation that can be implemented by adding an ancillary system and then forgetting it. This transformation can be chosen in many ways. I focussed on two particular examples, see Eqs. (\ref{V}) and (\ref{W}). The first one works for two systems of the same dimension and supports entanglement in the phase-locked subspace, whereas the second one is designed to observe synchronization of a single-qubit dynamics to a d-level stimulus. My results provide a supporting argument for single-qubit synchronization \cite{Qsync1,Qsync2,Qsync3,Qsync4}.

It is natural to investigate what possibilities other choices of the phase-locking transformation can offer. This might be particularly interesting in the context of quantum state-engineering and quantum control. Furthermore, research on the problems discussed in this work are relevant from the point of view of complex quantum dynamics, quantum chaos and pseudochaos. Note, that phase-locking in qudit systems can be considered as emergence of some order out of pseudochaos. More precisely, if the dimension of the system and the initial frequencies of the oscillator and the stimulus are co-prime, the initial dynamics is ergodic in the sense that it visits all states in the state-space. However, phase-locking brakes this ergodicity and the system evolves from the $d^2$-dimensional Hilbert space to a highly correlated $d$-dimensional attractor subspace.


{\it Acknowledgements.} I would like to thank Andy Chia and Karol \.{Z}yczkowski for helpful discussions. This work is supported by the Ministry of Science and Higher Education in Poland (science funding scheme 2016-2017 project no. 0415/IP3/2016/74). 



\end{document}